# Quantifying Atom-scale Dopant Movement and Electrical Activation in Si:P Monolayers

Xiqiao Wang,[1,2] Joseph A. Hagmann,[1] Pradeep Namboodiri,[1] Jonathan Wyrick,[1] Kai Li,[1] Roy E. Murray,[1] Alline Myers,[1] Frederick Misenkosen,[1] M. D. Stewart, Jr.,[1] Curt A. Richter,[1] Richard M. Silver[1]

[1] National Institute of Standards and Technology, 100 Bureau Dr., Gaithersburg, Maryland 20899, USA

[2] Chemical Physics Program, University of Maryland, College Park, Maryland 20742, USA

# Abstract

Advanced hydrogen lithography techniques and low-temperature epitaxial overgrowth enable patterning of highly phosphorus-doped silicon (Si:P) monolayers (ML) with atomic precision. This approach to device fabrication has made Si:P monolayer systems a testbed for multiqubit quantum computing architectures and atomically precise 2-D superlattice designs whose behaviors are directly tied to the deterministic placement of single dopants. However, dopant segregation, diffusion, surface roughening, and defect formation during the encapsulation overgrowth introduce large uncertainties to the exact dopant placement and activation ratio. In this study, we develop a unique method by combining dopant segregation/diffusion models with sputter profiling simulation to monitor and control, at the atomic scale, dopant movement using room-temperature grown locking layers (LL). We explore the impact of LL growth rate, thickness, rapid thermal anneal, surface accumulation, and growth front roughness on dopant confinement, local crystalline quality, and electrical activation within Si:P 2-D systems We demonstrate that dopant movement can be more efficiently suppressed by increasing the LL growth rate than by increasing LL thickness. We find that the dopant segregation length can be suppressed below a single Si lattice constant by increasing LL growth rates at room temperature while maintaining epitaxy. Although dopant diffusivity within the LL is found to remain high (on the order of $10^{-17} cm^2/s$) even below the hydrogen desorption temperature, we demonstrate that exceptionally sharp dopant confinement with high electrical quality within Si:P monolayers can be achieved by combining a high LL growth rate with a low-temperature LL rapid thermal anneal. The method developed in this study provides a key tool for 2-D fabrication techniques that require precise dopant placement to suppress, quantify, and predict a single dopant's movement at the atomic scale.

Highly phosphorus-doped silicon (Si:P) monolayers are a novel 2-D system that can be patterned with atomic scale precision and features high carrier densities.[1-3] They have attracted an enormous amount of interest with their potential applications in multiqubit quantum computers and atomically precise 2-D superlattice designs.[2, 4, 5] Advanced hydrogen lithography techniques and low-temperature epitaxial overgrowth enable individual dopant placement into Si lattice sites with atomic precision in all three dimensions.[6] In this way, atomically precise Si:P planar architectures, such as atomically abrupt wires,[7, 8] tunnel junctions,[9] quantum dots,[10, 11] single atom transistors,[2] and ordered single dopant arrays[5, 12] have been successfully defined on H-terminated Si(100) surfaces. These patterned devices are then encapsulated in epitaxial overgrown crystalline Si. Central to the fabrication and performance of these planar Si:P devices is the preservation of exact lattice locations of deterministically placed dopant atoms during overgrowth. In atomically precise few dopant quantum devices and superlattice dopant arrays, spatial fluctuations in dopant positions by even a single lattice constant can disrupt the quantum device performance and dramatically alter the quantum coupling.[13] In Si:P planar contact and gate regions, deviation of the 2-D dopant confinement from an ideal Si:P monolayer has profound effects on 2-D electrical properties .[14] Atomically sharp dopant confinement, high dopant activation ratios, and a defect-free epitaxial environment are essential attributes of proposed donor-based Si:P quantum computer architectures,[6, 11, 15] necessitating the development of precision metrological and fabrication methodologies to control dopant confinement and epitaxial quality at the atomic scale.[16] In this study, we develop a robust quantification method to monitor and control, at the ultimate monoatomic layer scale, unintentional dopant movement and formation of lattice defects to enable characterization and optimization of Si:P monolayer fabrication, fundamental to donor-based Si quantum computing and atomically precise 2-D superlattice design.

Encapsulation of a Si:P monolayer device within a crystalline Si matrix fully activates P dopants, isolates the conducting channels from the complex surface and interface interactions, and protects the Si:P system against ambient degradation.[17] However, dopant segregation, diffusion, and surface roughening during the epitaxial encapsulation process redistributes dopant atoms and introduces large positional uncertainties in the resulting dopant locations. [20, 21, 23, 30, 31] Defect formation in epitaxial Si overgrowth can create deactivation centers,[18] decrease free carrier mobility,[3] and increase noise floors in Si:P 2-D systems.[19] A key development to address the well-known trade-off between low-temperature encapsulation for sharp dopant confinement and high-temperature encapsulation for optimum epitaxial quality [20-22] has been the recent application of thin room-temperature grown layers, commonly referred to as locking layers (LL), followed by encapsulation overgrowth at elevated temperatures. [23-25] While theoretical calculations have been carried out on the effects of various levels of dopant confinement on Si:P 2-D properties,[1, 14] experimental quantification of dopant confinement and redistribution within room-temperature grown LLs remains challenging with little success at the monoatomic layer scale. The importance of this challenge is paramount to the development and performance of atomically precise 2-D superlattice designs and donor-based quantum computing.[2, 5]

In this study, we develop for the first time a robust method to quantify dopant movement at the atomic scale during Si:P monolayer fabrication by combining segregation/diffusion models with sputtering profiling simulations. Dopant segregation, diffusion, surface accumulation, and growth front roughening have been taken into account in this quantitative investigation on the impact of LL growth parameters on dopant confinement, local crystalline quality, and dopant activation in Si:P 2-D systems. The extraordinarily high dopant density within the 2-D layers and the kinetically controlled 3-D island growth front during the room temperature LL overgrowth create a complex yet unique 2-D system environment that has been studied little to date. We experimentally determine, for the first time, the LL growth rate dependence of the dopant segregation length and the dopant diffusivity within LLs below the hydrogen desorption temperature. We combine Scanning Tunneling Microscopy (STM), Transmission Electron Microscopy (TEM), Secondary Ion Mass Spectroscopy (SIMS), Atom Probe Tomography (APT), and low-temperature magnetotransport measurements to obtain detailed insight into optimizing Si:P 2-D system fabrication at the individual atom layer scale. The locking layer overgrowth parameter space explored in this study is fully compatible with current state-of-the-art hydrogen lithography techniques and can be applied directly to fabricate atomically precise superlattices and quantum devices.

## Methods

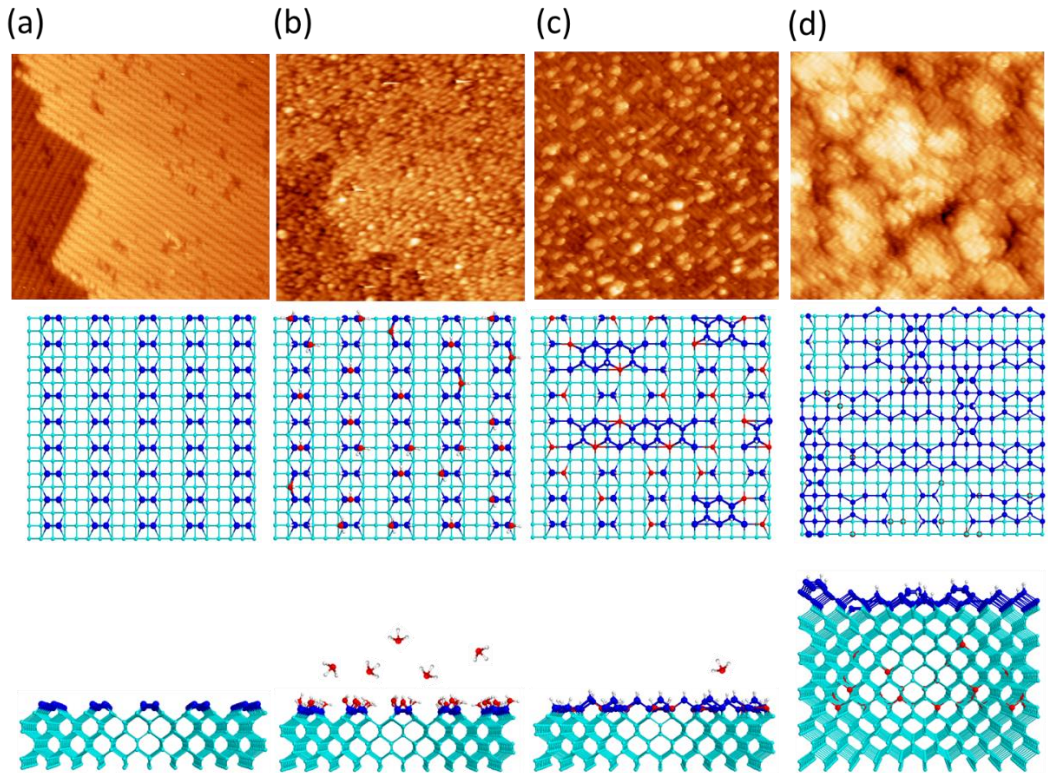

Figure 1. STM topography images (+2V bias on substrate, 0.2nA set-point current, 25nm × 25nm, acquired from different samples at different stages of preparation) with complementary atomic lattice top and side view schematics of the phosphine dosing, incorporation, and encapsulation processes on a blanket Si(100) 2×1 surface. In the schematic figures, the blue and cyan atoms represent Si on the surface and in bulk, respectively. Red atoms represent P, and white atoms represent H. (a) A typical starting Si(100) surface with a 2×1-dimer row reconstruction and the characteristic alternating dimer rows across a step edge. (b) The Si(100) surface covered with ~0.37 monolayers of adsorbed $PH_x$ (x=0,1,2) groups after saturation dosing (approximately 1.5 Langmuir exposure) at room temperature. (c) The surface after an incorporation flash anneal with the brighter regions being islands formed by ejected (substituted) Si atoms. Since the ejected Si should be in one to one correspondence with incorporated P atoms, the ejected Si island coverage represents the incorporated P concentration.[2, 26, 27] (d) The growth front morphology of a nominal 274°C overgrowth at on top of the P-incorporated surface. The overgrowth is in the kinetically rough growth mode due to limited Si adatom migration on the growth front. Though it is difficult to distinguish P atoms on a rough growth front,[28] as shown in the side view schematics (bottom panel), the incorporated P atoms segregate above the original doping plane during the 274°C overgrowth, which broadens the delta layer.

Si:P monolayers are fabricated using atomic layer doping.[29-31] Figure 1 illustrates the Si:P 2-D system fabrication process. The samples discussed in this study were fabricated on 1-10 ohm-cm boron doped p-type Si chips. First, an atomically flat, clean Si(100) 2×1 reconstructed surface is prepared in an ultrahigh vacuum (UHV) system with a $6.6 \times 10^{-9}$ Pascal ($5 \times 10^{-11}$ Torr) base pressure, Figure 1(a). Detailed preparation procedures have been published elsewhere.[32] Then the surface is dosed (~1.5 Langmuir exposure) with Phosphine ($PH_3$) gas at room temperature to achieve a saturation surface coverage of ~0.37 monolayers of phosphorus species (Figure 1(b)).[22, 33] $PH_3$ molecules dissociate into H atoms and $PH_x$ (x= 0, 1, 2) groups and terminate the Si dangling bonds on the Si(100) surface.[34, 35] A Rapid Thermal Anneal (RTA) at nominally 384°C for 2 min incorporates the P atoms substitutionally into the silicon lattice within the first atomic layer.[36-38] This P incorporation enhances the electrical activation of the dopants and helps minimize segregation during the subsequent Si overgrowth process.[36] The substituted Si atoms in the top layer eject onto the surface and form short 1D Si chains perpendicular to the underlying dimer rows, Figure 1(c).[26, 27] Some of the Si surface bonds are terminated by H atoms that dissociate from phosphine molecules. This phosphorus incorporation process results in a partially hydrogen-terminated Si(100) surface with approximately one quarter to one third monolayer coverage [22, 33, 35, 36] of incorporated P atoms.

The SIMS measured P concentration is $(2.0 \pm 0.2) \times 10^{14}/cm^2$ in our delta layer samples after encapsulation overgrowth is consistent with the ejected Si atom coverage. The side-view schematics in Figure 1(d) demonstrate P segregation during low-temperature encapsulation, which results in P moving away from the original doping plane, broadening the confinement of P atoms asymmetrically in the overgrowth direction. It is well known that temperature measurement of silicon in the low-temperature range ( below ~400°C) and in a UHV environment is challenging and is likely to be the largest source of chamber-to-chamber variation in low-temperature epitaxial growth.[28, 39-41] In this study, sample temperatures are measured using infrared pyrometers with the emissivity value calibrated using Au-Si (363°C, 97.15/2.85 wt-%) eutectic alloys on Si substrates in a high vacuum environment. The encapsulation overgrowth temperature and locking layer rapid thermal anneal temperature are 274±0.2°C and 384±0.2°C respectively, where the uncertainties are given as one-sigma standard deviations. We overgrow Si using a Silicon Sublimation Source (SUSI-40) by passing DC current through a high-purity intrinsic Si filament,[42] which is shielded by Si from any hot metal and ceramic components to prevent contamination. The SUSI growth rate is calibrated by using phase shift interferometry, SIMS, cross-section TEM results as well as imaging sub-monolayer deposition using STM. The calibrated SUSI growth rate has been published elsewhere.[42] The SIMS measurement of the P concentration profile uses a Cs+ primary ion-beam with an acceleration energy of 1 keV or 0.3keV and an incident angle of 60°. Negative ions of 30Si+31P are measured to obtain phosphorus concentration profiles. The estimated calibration uncertainty for P quantification is nominally ± 10%.

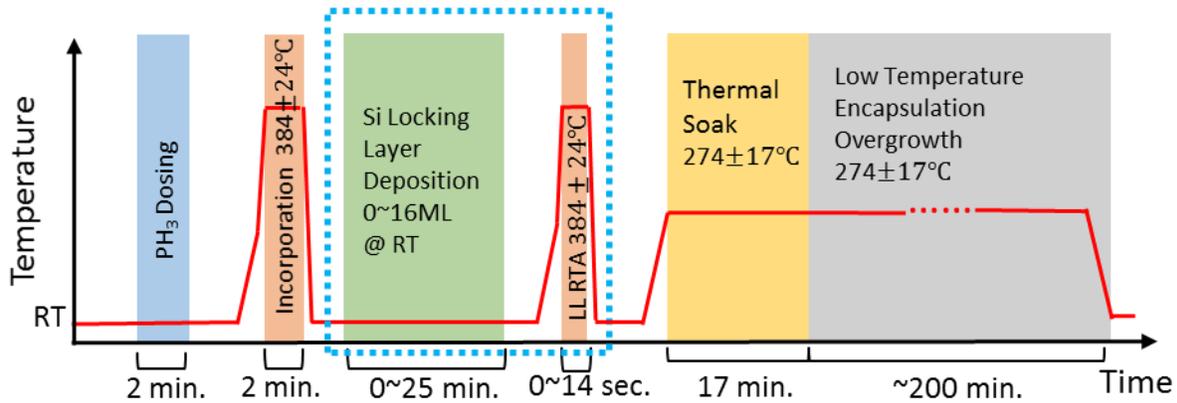

Figure 2. The process flow diagram of the delta layer fabrication procedures illustrating the timing and temperature at each step of the process. The blue box highlights the steps that were systematically varied in this study: the locking layer (LL) overgrowth varies from 0 ML to 16 ML with or without a subsequent LL Rapid Thermal Anneal (RTA) at 384°C for 14 sec. The red line represents the thermal profile as a function of time.

An optimized locking layer (LL) deposited at room temperature followed by encapsulation overgrowth at elevated temperatures is critical to simultaneously suppress dopant segregation and maximize crystalline quality at the Si:P 2-D system.[23, 24, 43] The maximum epitaxial thickness, beyond which overgrowth becomes amorphous, decreases rapidly at reduced temperatures due to surface roughening.[39, 44] On Si(100) surfaces, the limiting epitaxial thickness falls below 3nm for room temperature overgrowth, which is insufficient to isolate the 2-D Si:P system from interface states and traps.[17] The essential idea behind LL overgrowth is that dopant segregation can be greatly suppressed during room-temperature LL overgrowth. Before reaching the limiting epitaxial thickness for room-temperature growth, the overgrowth temperature is increased to sustain the epitaxial growth mode.[23, 25, 43] Figure 2 illustrates the entire growth process for a Si:P monolayer, locking layer (LL), and encapsulation overgrowth. Before starting the low-temperature encapsulation at 274°C, the sample temperature is maintained for 17 min to stabilize the temperature and Si deposition rate.[45] As a result, the surface undergoes a low temperature thermal anneal before each deposition step at elevated temperatures. We will discuss the effect of this pre-deposition anneal on the LL in a later section.

## Results and discussions

### Epitaxial quality at locking layer interface

(a)

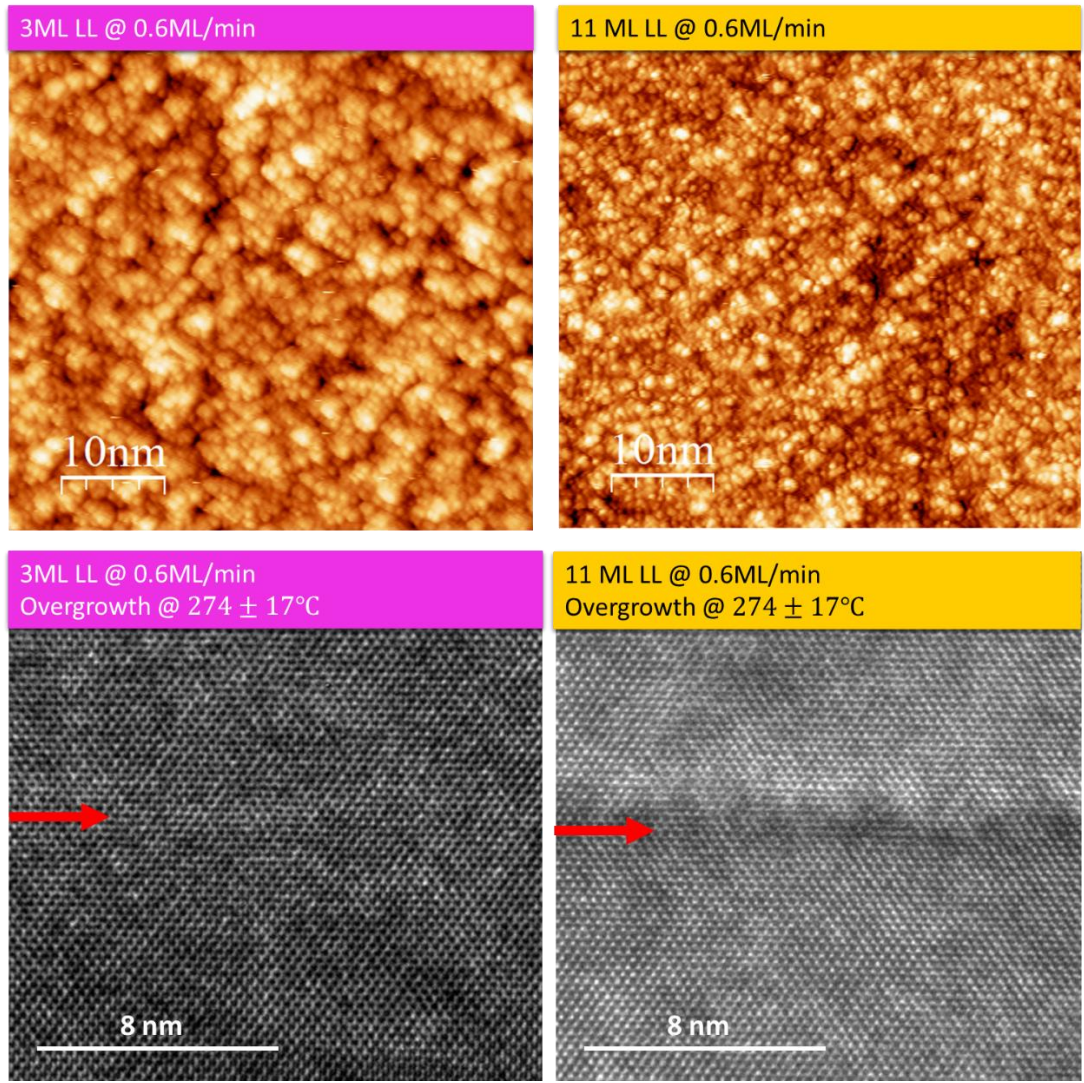

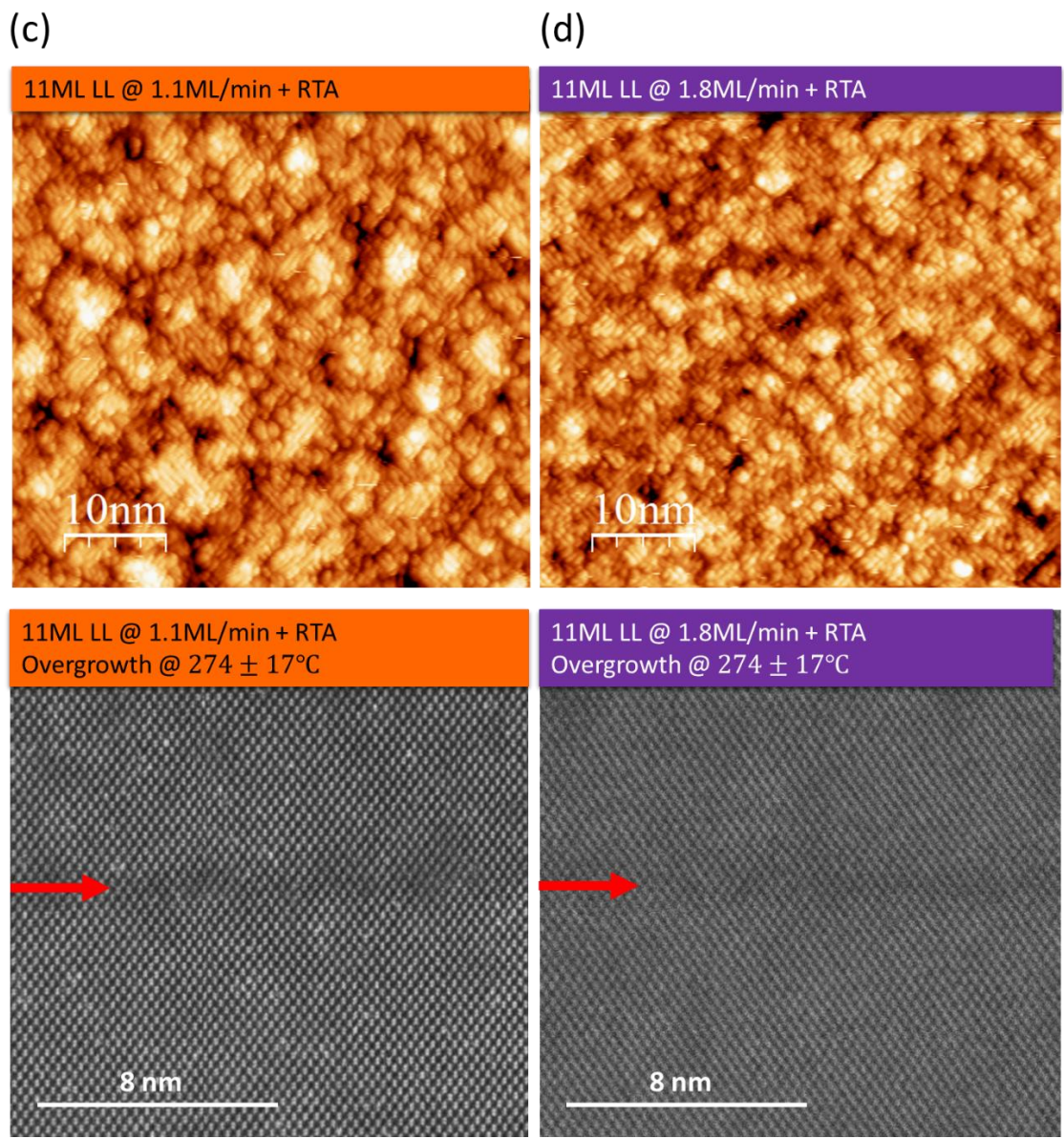

Figure 3. Top panels: STM topography images (+2V bias on sample, 0.2nA set-point current) of various LL surfaces before low temperature encapsulation. Bottom panels: High-resolution cross section TEM/STEM micrographs near the LL interface regions after LL deposition and low temperature encapsulation overgrowth. The locking layer growth conditions (thickness, growth rate, and rapid thermal anneal (RTA)) and the subsequent encapsulation overgrowth are marked in the graphs. The red arrows in TEM/STEM images indicate the LL interfaces.

STM micrographs of LL surface morphology prior to low temperature encapsulation overgrowth are shown in the top panels in Figure 3. Compared with the surfaces after P incorporation (Figure 1(c)), the LL deposition introduces high island/step densities on the low temperature encapsulation overgrowth starting surface. The bottom panels in Figure 3 show high-resolution cross-section TEM/STEM micrographs near the locking layer interface regions after LL deposition and low temperature encapsulation overgrowth. The lattice planes align very well across the doping plane, and no distinction in crystalline quality can be observed between the encapsulation overgrowth layers and the substrates, indicating good epitaxial overgrowth quality in the encapsulation layer grown at 274°C. Thin (3ML) LL deposition on top of Si:P monolayer at room temperature is within the kinetically controlled 3-D island growth mode as a result of negligible Si adatom surface migration (Figure 3(a)).[46, 47] We observe no interface contrast at the 3ML LL plane, which indicates that excellent epitaxial quality can be maintained at a few-ML RT-grown LL interface. Thicker RT-grown LLs lead to smaller 3-D island sizes and higher LL surface roughness (Figure 3(b)), which may affect the epitaxial quality within the LL and alter the initial surface conditions for subsequent low-temperature encapsulation overgrowth.[48] In contrast to Figure 3(a), faint dark TEM contrast is observable at the thicker (11ML) LL interface in Figure 3(b), which is likely caused by a higher concentration of defects and increased strain at the thicker LL interface region. However, the detailed physical mechanism at the thicker LL interface remains to be explained. Annealing at elevated temperatures is known to repair Si lattice defects and interstitial dopant defects and decreases local lattice strain.[49] In Figure 3(c), an RTA at 384°C for 14 seconds flattens the LL surface and improves the LL crystallinity because of an increase in island size and diffusion of Si atoms to step edges. regardless of the higher growth rate.[32] The surface roughness effect from higher locking layer growth rates are not obvious after LL RTA (Figure 3(d)). However, TEM contrast at the LL interface (Figure 3(c)(d)) remains observable after such a short RTA process.

## *Modeling the P-profile with locking layers*

The depth resolution of the SIMS technique is on the order of several nanometers due to atomic mixing and sputter roughening effects during the profiling process. It has been recognized that some correction to the measured SIMS data, which takes into account distortion effects from the sputtering process, is necessary to obtain the true composition depth profile from the measurement.[50-53] The measured SIMS profile is a convolution of the real P concentration profile with a sputtering depth resolution function. Quantifying the concentration profile with sub-nanometer depth resolution can only be accomplished by applying an appropriate deconvolution or through profile reconstruction methods.[54] A direct deconvolution is complicated and yields large errors due to measurement signal noise.[50, 51, 55] In this study, we fit a simulated convolution to the measured SIMS results and reconstruct the actual dopant concentration profile using the best-fit parameters. We use a first order segregation model to simulate the dopant concentration profile. A second order segregation component is unnecessary because the P coverage on the growth front surface of this study is not high enough to form P-P donor pair defects,[56-58] which is considered the primary cause of the breakdown of the first order model.[59] The depth resolution function is simulated using the Mixing-Roughness-Information-depth (MRI) sputter profiling model[50-52, 54, 55, 60] to account for sputtering-induced broadening effects during the SIMS measurement.

Dopant segregation during epitaxial Si overgrowth is such that as a new monolayer overgrows on top of the surface, a portion of the P atoms on the initial surface float onto the new surface due to the lower configuration energy on the surface (segregation energy).[59, 61, 62] This segregation proportion depends critically on overgrowth temperature, overgrowth rate, and the initial surface conditions such as surface step density and surface passivation conditions.[56] In our first order segregation model, the total overgrowth is divided into a LL region and an encapsulation region. A constant incorporation probability $a_{LL}$ in the LL region ($a_{CAP}$ in the encapsulation region) is defined as the percentage of the surface phosphorus atoms that are incorporated into the existing layer as another monolayer of Si atoms is overgrown on top of that layer. The segregation model is expressed in the following form,

$$\frac{dN_{surf}}{dx} = a_i N_{surf} \qquad \text{(Equation 1.)}$$

where $N_{surf}$ is the phosphorus atom density on the growth front surface; $x$ is overgrowth thickness in units of ML. The segregation length in each region, $l_i$ is defined as the length for the 1/e monolayer coverage decrease. It follows that $l_i = \frac{1}{a_i}$.

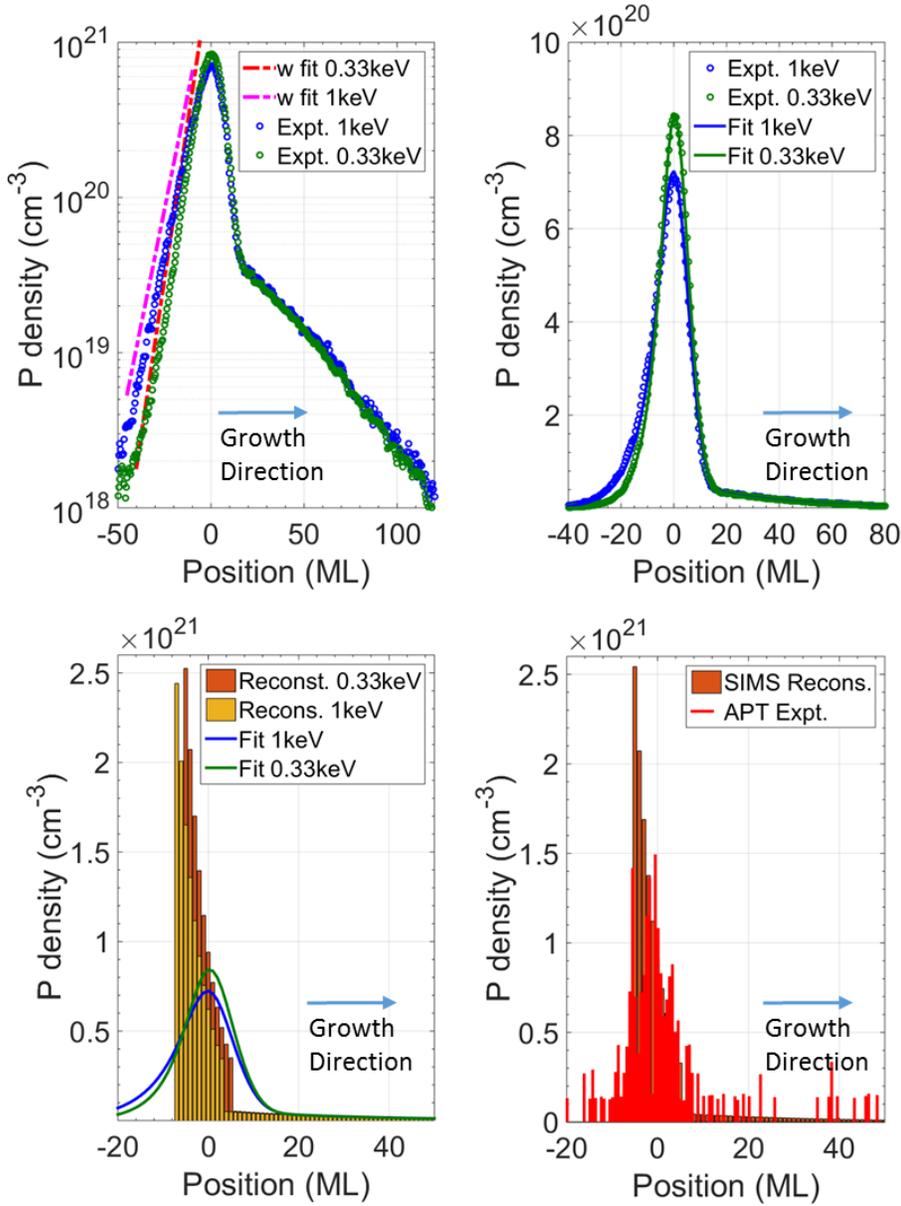

Figure 4. Reconstruction of the physical dopant concentration profiles from SIMS measurements. 1keV and 0.3keV primary ion beam energies are used for SIMS measurements on the individual LL sample (see Sample LL-T3 in Table 1.). (a) The atomic mixing length ($w$) depends critically on the primary ion beam energy and is obtained by fitting the trailing edge of the measured SIMS profile M(x). (The fitted $w$ lines are shifted to avoid masking the data points) (b) The SIMS data and the fitted SIMS results M(x) are plotted as data points and solid curves. We intentionally shift the zero position of the measured SIMS profile peaks for comparison purposes. (c) the reconstructed concentration depth profiles N(x) are plotted in bars. Each bar represents 1ML. (d) Comparison between the reconstructed P concentration profile N(x) and the atom probe tomography (APT) result.

The MRI sputter profiling convolution is governed by three well-defined physical parameters: the atomic mixing length $w$, the roughness $\sigma$, and the information depth $\lambda$. The atomic mixing length $w$ depends critically on the sputtering primary ion beam energy and is obtained by fitting the exponential section of the trailing edge of the profile (Figure 4(b)).[63] The roughness, $\sigma$, consists of contributions from the surface roughness of the original dosing plane due to steps and kinks, the surface roughness after overgrowth, sputtering induced surface topography, and mixing length straggling.[50, 64] The information depth $\lambda$ for SIMS is given by the escape depth of the secondary ions. Since the sputtered secondary ions are from the top layer in SIMS measurements with low primary ion beam energies, we take $\lambda$ to be 1 ML in this work.[50]

First, the physical bulk concentration profile $N(x)$ is obtained by calculating the surface concentration as the overgrowth proceeds layer by layer using the recurrence relation implied by Equation 1. We emphasize that $N(x)$ represents the physical bulk concentration assuming an atomically flat single terrace initial dosing plane. Atomic layer steps and kinks could introduce surface roughness on the initial dosing plane. In this study, the initial dosing plane roughness is included in the total roughness parameter, $\sigma$, which is to be convoluted with $N(x)$ in the next step. Recently, our group has shown that a large atomically flat single terrace dosing plane can be formed on micropatterned Si(100) in a controlled way,[65] where the reconstructed profile $N(x)$ will represent the real physical bulk concentration at local single terrace regions. In the next step, the three convolution functions, $g_w$, $g_\sigma$, and $g_\lambda$, are sequentially applied to $N(x)$ to obtain the sputter convoluted profile $M(x)$, as shown in Equation 2. $w_0$ and $\lambda_0$ are the respective normalization factors of $g_w$ and $g_\lambda$ for the conservation of the total number of phosphorus atoms. The total concentrations of P atoms are obtained by integrating the SIMS depth profiles and used as an input normalization factor. Since the segregation length in the low-temperature encapsulation overgrowth layer is much longer than the characteristic sputtering length scales ($w, \sigma, \lambda$), we obtain $a_{CAP}$ by directly fitting the exponential section of the leading edge of the encapsulation layer profile above the LL. By using the pre-fitted $w$ and $a_{CAP}$ as inputs, the LL incorporation probability ($a_{LL}$) and surface roughness ($\sigma$) are treated as independent fitting parameters to fit $M(x)$ to the measured SIMS profiles.

$$M(x) = \int_{-\infty}^{+\infty} N(x')g(x-x')dx' \qquad \text{(Equation 2.)}$$

$$g(x) = g_w(x) * g_\sigma(x) * g_\lambda(x)$$

$$g_w(x) = \begin{cases} \frac{1}{w_0}\exp\left[\frac{-(x+w)}{w}\right] & x > -w \\ 0 & x \leq -w \end{cases}$$

$$g_\sigma(x) = \frac{1}{\sigma\sqrt{2\pi}} \exp\left[\frac{-x^2}{2\sigma^2}\right]$$

$$g_\lambda(x) = \begin{cases} \frac{1}{\lambda_0} \exp\left(\frac{x}{\lambda}\right) & x \leq 0 \\ 0 & x > 0 \end{cases}$$

In Figure 4, we numerically fit two SIMS profiles measured on the same delta layer sample but with different primary ion beam energies of 1keV and 0.3keV. The depth is in units of monolayer (ML) thickness, and the SIMS-measured concentration peak positions are shifted to the zero-depth position for comparison. When fitting the depth profile, data points are weighted by the deviation of their Poisson error. The individual fitting parameters $a_{LL}$ and $\sigma$ are only weakly correlated with each other (Pearson correlation coefficient $< 0.5$ between $a_{LL}$ and $\sigma$). As can be seen from the best-fit parameters in Table 1, a 0.3keV beam energy results in a smaller atomic mixing length (~5.4 ML) than the 1keV beam energy does (~7.3ML). The simulation separates the sputter broadening effects from the actual P-profile and the reconstructed profiles at 1keV and 0.3keV show excellent agreement with each other, independent of sputter beam energies. As can be seen from Table 1, the segregation incorporation probabilities during 274°C overgrowth ($a_{CAP}$) are approximately one order of magnitude lower than that during RT LL overgrowth ($a_{LL}$), which accounts for the concentration discontinuity between the LL and subsequent encapsulation overgrowth layer. The best-fit sputtering front roughness ranges approximately from 3 to 4 ML for samples with a LL, which is in good agreement with the observed surface roughness in AFM and cross sectional TEM images. As shown in Figure 4(c), due to the atomic mixing effect, the measured SIMS concentration peaks lie shallower than the reconstructed profile peaks. The dependence of the measured SIMS profile peak positions on the sputter ion beam energy highlights the importance of using profiling reconstruction techniques to extract the real depth information of incorporated dopant atoms following atomic device encapsulation.

As shown in Figure 4(d), our reconstructed profile agrees well with the Pulsed Laser Atom Probe Tomography (PLAPT) measurement result. We note that several factors can limit the resolution of the APT technique, such as low counting number noise, the evolution and local variation of tip shape, field induced surface migration, crystallographic dependence of evaporation fields between Si and P species, and aberration effects, *etc*.[66-69] A detailed comparison between the SIMS reconstruction and APT reconstruction techniques at the ultimate monolayer limit will be published elsewhere.

Table 1 summarizes the detailed LL fabrication parameters (LL thickness, LL growth rate, and LL RTA) of LL samples investigated in this study as well as the best-fit parameters. Uncertainties are given as one-sigma standard deviations, which include only statistical uncertainties.

| Sample Name | LL Thickness (ML) $\sigma \leq 15\%$ | LL Growth Rate (ML/min) $\sigma \leq 15\%$ | LL RTA (384°C 14sec.) | Primary beam energy (keV) | $a_{LL}$ (/ML) $\sigma \leq 20\%$ | $a_{CAP}$ (/ML) $\sigma \leq 20\%$ | $D\ (cm^2/s)$ $\sigma \leq 50\%$ | Mixing Length $\omega$(ML) $\sigma \leq 10\%$ | Roughness (ML) $\sigma \leq 20\%$ |
|---|---|---|---|---|---|---|---|---|---|
| LL-T0 | 0 | -- | No | 1.0 | -- | 0.018 | -- | 7.3 | 4.2 |
| LL-T1 | 3 | 0.6 | No | 1.0 | 0.18 | 0.043 | -- | 7.4 | 4.1 |
| LL-T2 | 6 | 0.6 | No | 1.0 | 0.18 | 0.041 | -- | 9.5 | 2.8 |
| **LL-T3** | 11 | 0.6 | No | 1.0 | 0.18 | 0.032 | -- | 7.3 | 3.2 |
| **LL-T3** | 11 | 0.6 | No | 0.3 | 0.19 | 0.033 | -- | 5.4 | 3.2 |
| **LL-T4** | 16 | 0.6 | No | 1.0 | 0.20 | 0.046 | -- | 7.7 | 3.3 |
| **LL-R1** | 11 | 1.1 | No | 1.0 | 0.24 | 0.016 | -- | 8.1 | 2.1 |
| **LL-R1-RTA** | 11 | 1.1 | Yes | 1.0 | 0.23 | 0.025 | $3.2 \times 10^{-17}$ | 7.7 | 2.8 |
| **LL-R2-RTA** | 11 | 1.8 | Yes | 1.0 | 0.29 | 0.046 | $1.3 \times 10^{-17}$ | 7.1 | 3.0 |

Table 1. Summary of the LL fabrication parameters (LL thickness, LL growth rate, and LL RTA) of LL samples investigated in this study and the best-fit parameters from the P-profile modeling. Uncertainties are given as one-sigma standard deviations, which only include statistical uncertainties. $a_{LL}$ and $a_{CAP}$ are dopant segregation incorporation probabilities during the LL overgrowth and encapsulation overgrowth, which represent the probabilities that a dopant on the surface monolayer remains within the same layer and does not segregate onto the upper layer during the subsequent one monolayer overgrowth. w is the atomic mixing length in sputtering process. Roughness is the sputter milling front roughness that consists of contributions from both the original surface/interface roughness and the sputtering induced surface topography.

## *Locking Layer Thickness*

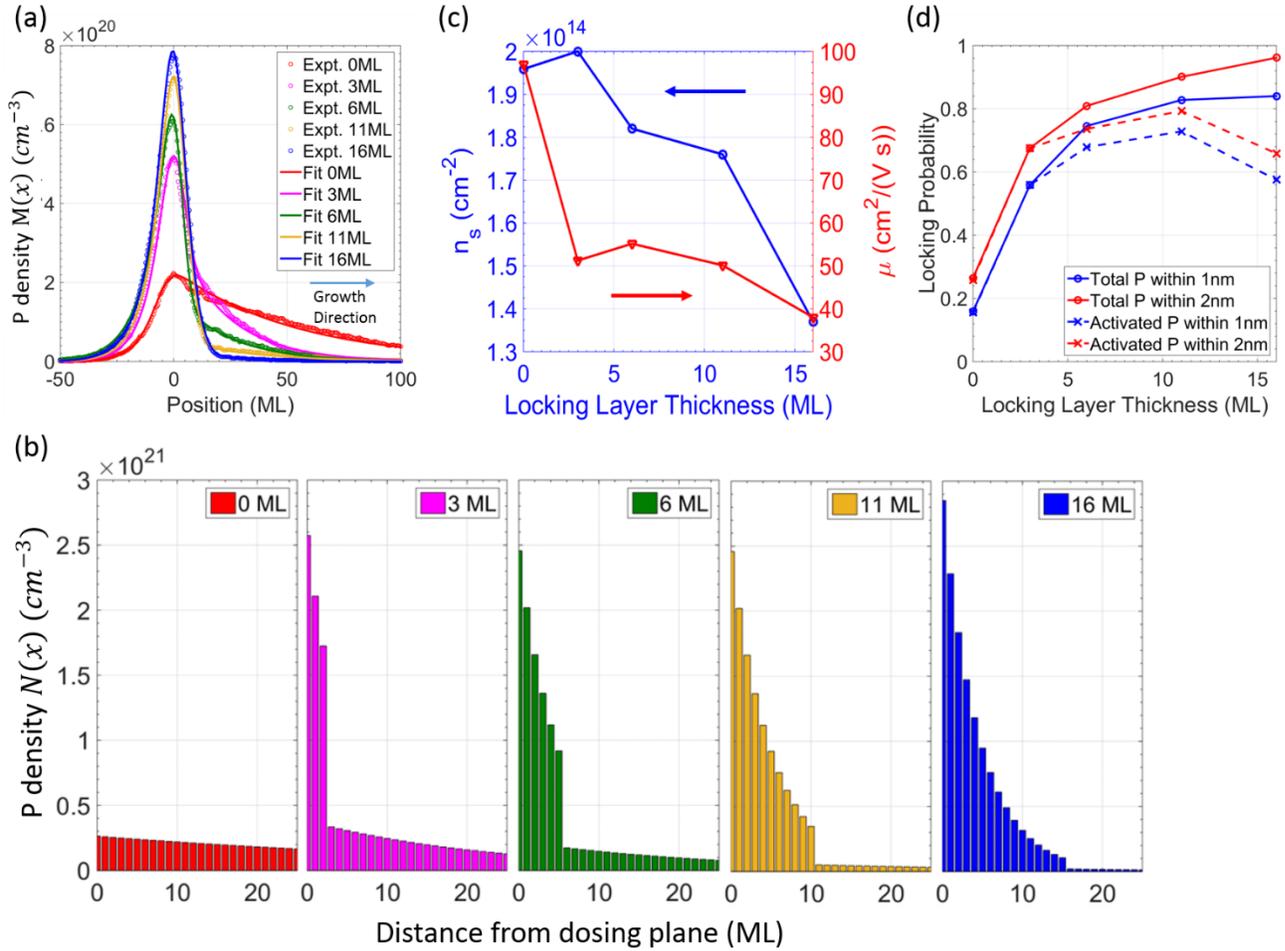

Figure 5. The effect of locking layer (LL) thickness on delta layer confinement and electrical properties. All locking layers are grown at 0.6ML/min at room temperature with no LL RTA. (a) The measured and fitted SIMS concentration profiles of LL samples with different LL thicknesses (see Samples LL-T0, LL-T1, LL-T2, LL-T3, and LL-T4 in Table 1). (b) The reconstructed P concentration profiles. (c) The delta layer free carrier mobility $\mu$ ($cm^2/(V\,s)$) and 2D sheet carrier density $n_s$ ($cm^{-2}$) are characterized at T=2 K using the van der Pauw technique. (d) The total and activated P locking probability 1nm and 2nm from the initial dosing plane as a function of LL thickness.

Figure 5 shows the effect of LL thickness on the delta layer confinement and electrical properties. All the LLs in Figure 5 are deposited at 0.6ML/min at room temperature without a LL RTA. Figure 5(b) illustrates the reconstructed P concentration profiles ($N(x)$) with different LL thicknesses. Without a LL, all the dosed P atoms in the initial dosing plane experience a high segregation probability with the encapsulation overgrowth at 274°C. Due to reduced segregation in the LL overgrowth, the rates at which P dopants are reduced with each ML overgrowth in LLs are much higher than those in encapsulation layers. At the same LL segregation probability, increasing the LL thickness drives down the remaining number of P atoms on the LL surface that experience a higher segregation probability in the subsequent 274°C encapsulation overgrowth as expected from equation 1. The reconstructed concentration profiles give approximately the same peak height at the dosing plane independent of LL thicknesses. It is the atomic mixing effect that accounts for the measured concentration peak height variations at different LL thicknesses in Figure 5(a).

Both the free carrier mobility and dopant activation ratio in the delta layers decrease as the LL thickness increases (Figure 5(c)). This drop in carrier density for samples with thicker LLs may be attributed to the formation of nonincorporated interstitial dopants, inactive dopant-vacancy complexes,[70] and deep level point defects in the lattice[39] as evidenced by the degradation in crystal quality (see Fig. 3(a, b)).[71] In Figure 5(d), we define the total locking probability as the probability for a single phosphorus atom to remain within a certain distance from the initial dosing plane after the entire encapsulation overgrowth process. The activated locking probability is calculated by multiplying the total locking probability with the dopant activation ratio. As expected, the total locking probability increases monotonically with LL thickness. However, the activated locking probability reaches its maximum at 11ML and decreases at 16ML LL thickness due to the inverse relationship between dopant confinement and activation ratio.

### *Locking Layer Rapid Thermal Anneal*

Keizer and coworkers have found that a finely tuned LL rapid thermal anneal (RTA) can effectively restore the active carrier density while maintaining ultra-sharp dopant profiles.[20, 56] They observed that application of a LL RTA slightly reduces the P peak height and raises the segregation tail of the encapsulation layer. We observe similar behavior in SIMS measured results when applying a short RTA (384°C for 14s) after RT LL overgrowth (Figure 6(a)(b)). This RTA induced dopant redistribution can be quantified by adding a diffusion component into our simulation algorithm to account for the P diffusion towards the surface during LL RTA (Equation 3), where the segregation profile after the RT LL deposition is used as the initial condition for the diffusion simulation. The diffusion equation is expressed as,

$$\frac{\partial N}{\partial t} = D \frac{\partial^2 N}{\partial x^2}$$   (Equation 3.)

Where $N$ is the phosphorus atom density in each monolayer, t is flash anneal time in seconds, $x$ is depth in units of ML, D is the diffusivity in units of $ML^2 \cdot s^{-1}$ and is treated as an independent fitting parameter. Since the RTA temperature (384°C) is well below the thermal desorption temperature of incorporated phosphorus atoms on Si(100) surfaces (≈600°C),[20, 33] we treat the phosphorus accumulation on a LL surface during a LL RTA as a diffusion sink where the diffusing P atoms remain trapped on the LL surface during an RTA. Dopant diffusion from surface into the overgrowth silicon is negligible within the low temperature range of this study because this process must overcome not only the diffusion barrier but also the segregation energy at the surface. Only the phosphorus atoms in the LL surface monolayer participate in the segregation process of the subsequent encapsulation overgrowth at 274°C.

Extrapolations from previous diffusivity studies show an over five orders of magnitude difference between the diffusivity of P in Si at our LL RTA temperature (384°C) and encapsulation temperature (274°C).[72-76] Therefore, we assume the P diffusion during the 274°C thermal soak and encapsulation overgrowth is negligibly small and is not included in our model.[24, 17, 57] Dopant diffusion into the substrate Si is also neglected at low temperatures in this study due to the low number of defects present in the Si substrate after flash annealing at 1200°C.[74, 77-79]

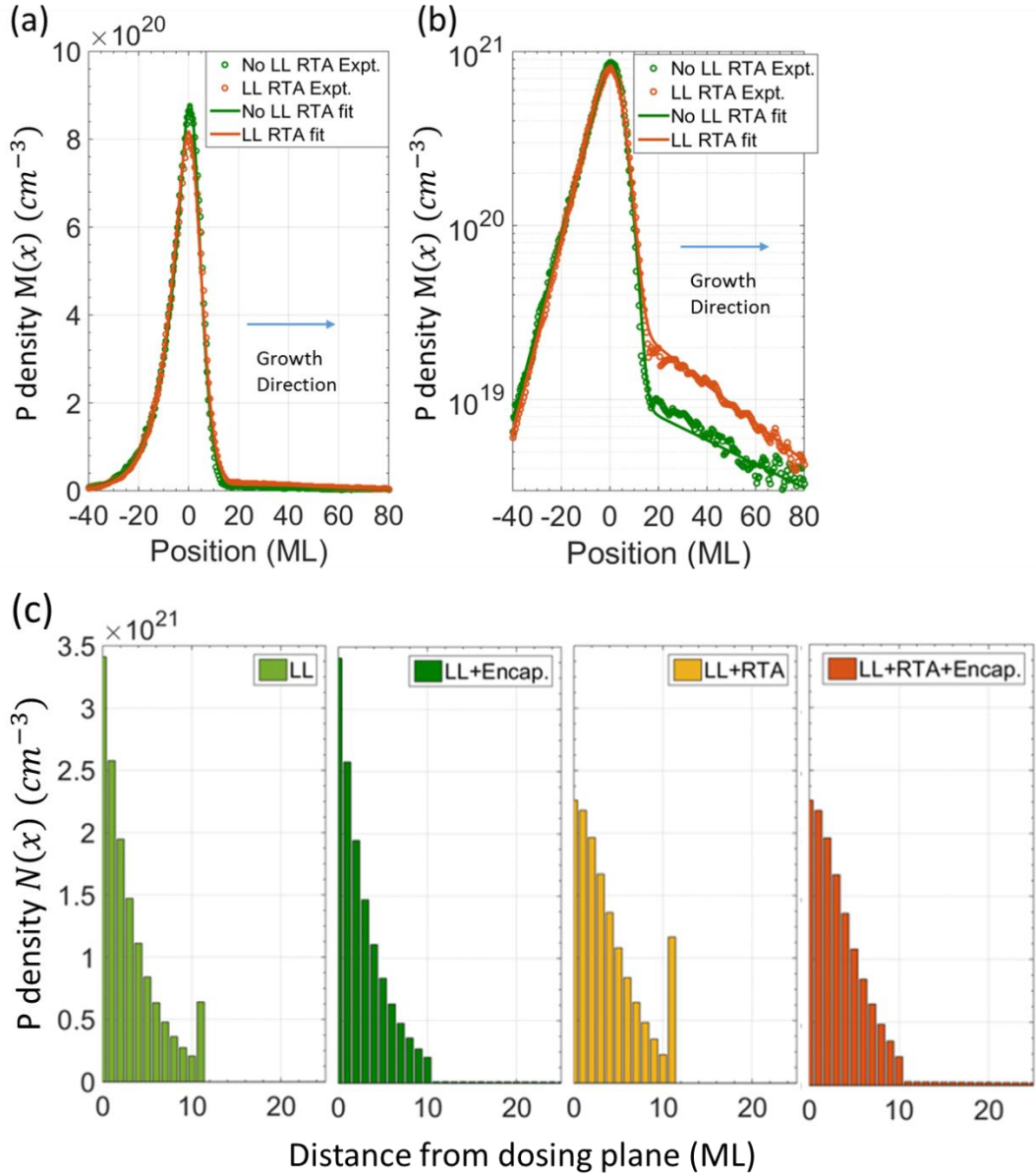

Figure 6. The locking layer (LL) rapid thermal anneal (RTA) effect on dopant redistribution in Samples LL-R1 and LL-R1-RTA. (a, b) The measured and fitted SIMS profiles. Sample LL-R1 has an 11ML LL grown at 1.1ML/min at room temperature without RTA. Sample LL-R1-RTA has the same RT-grown LL followed by a 384°C RTA for 14 seconds before low temperature encapsulation overgrowth. (c)The reconstructed P concentration profiles before and after low temperature encapsulation overgrowth in Sample LL-R1 (left two panels) and Sample LL-R1-RTA (right two panels).

We apply LL RTA to two of the samples in this study, Sample LL-R1-RTA and Sample LL-R2-RTA, where the LLs of the same thickness (11 ML) are grown at 1.1ML/min and 1.8ML/min respectively. We obtain the best-fit LL diffusivity to be about $3.2 \times 10^{-17} cm^2/s$ for Sample LL-R1-RTA and about $1.3 \times 10^{-17} cm^2/s$ for Sample LL-R2-RTA (see Table 1). Among the three free fitting parameter ($a_{LL}$, $\sigma$, and $D$), a relatively strong correlation exists between $a_{LL}$ and $D$ (Pearson correlation coefficients $\approx -0.2$ between $D$ and $\sigma$, and $\approx 0.9$ between $D$ and $a_{LL}$). However, the best-fit $a_{LL}$ value in Sample LL-R1-RTA shows good agreement with the best-fit $a_{LL}$ value in Sample LL-R1 where the RT-grown LL is deposited at the same deposition rate and thickness but without a LL RTA. This indicates that the simulation can distinguish the diffusion effect from the segregation effect in the SIMS profiles. As illustrated in the first and third panels of Figure 6(c), before low temperature encapsulation overgrowth, the LL RTA induces dopant atom diffusion within the LL, which reduces dopant density at the initial dosing plane and drives some subsurface dopant atoms out of the LL to the surface. This dopant accumulation on the LL surface results in slightly higher dopant concentration in the subsequent encapsulation overgrowth layer because the subsequent segregation starts with a higher initial surface coverage (second and fourth panels in Figure 6(c)).

### *Locking Layer Growth Rate*

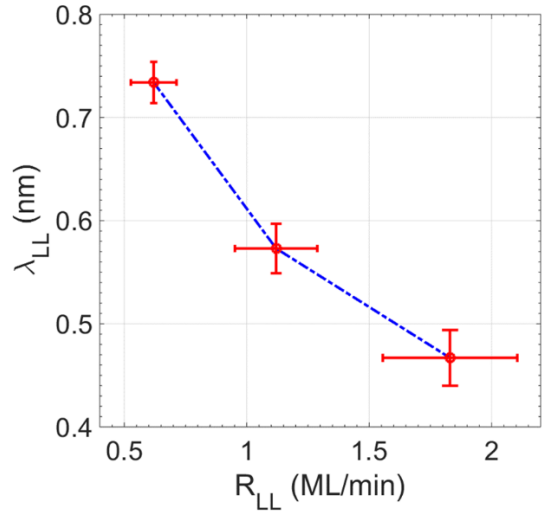

Figure 7. The fitted P segregation length ($l_{LL}$) of room-temperature grown locking layer as different growth rates.

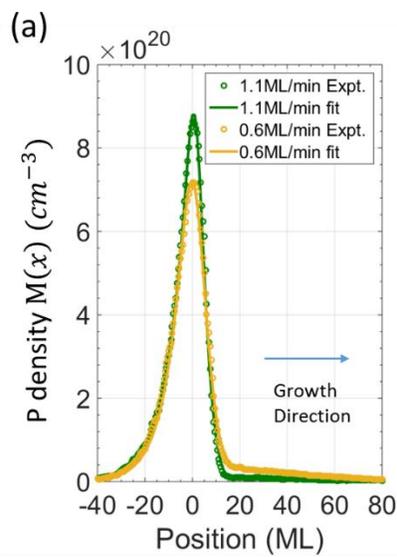

(a)

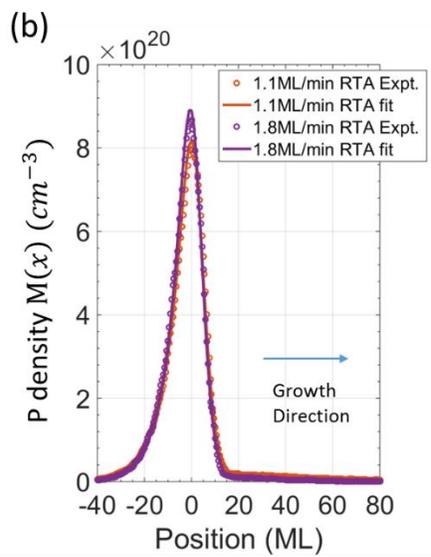

(b)

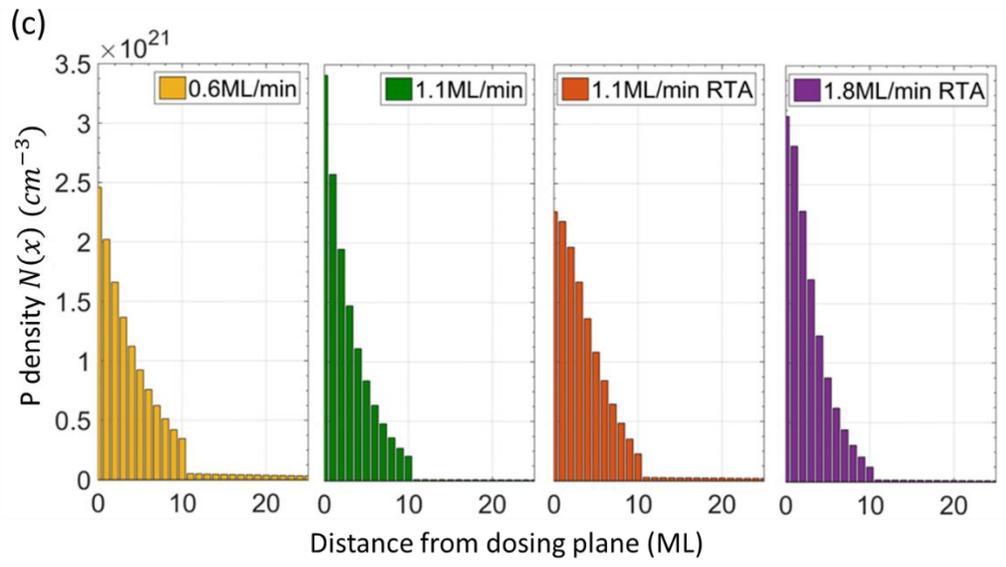

(c)

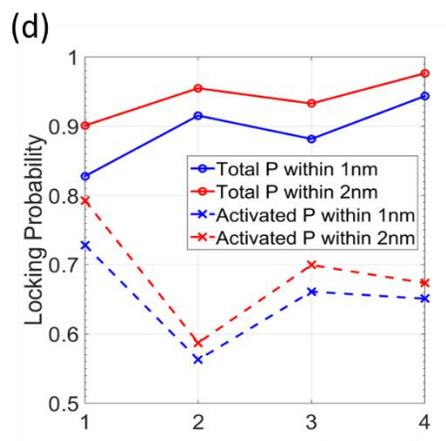

(d)

(1) 0.6ML/min    (3) 1.1ML/min + RTA
(2) 1.1ML/min    (4) 1.8ML/min +RTA

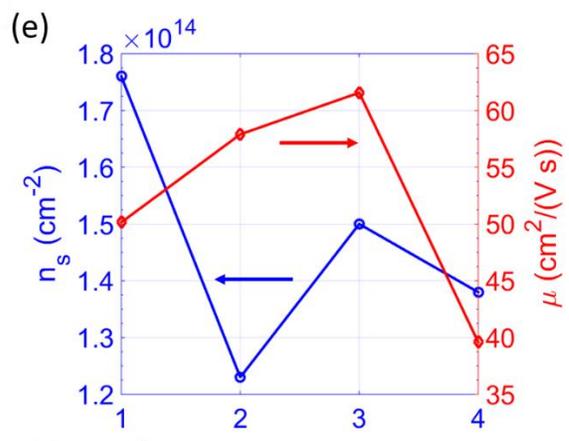

(e)

Figure 8. The effect of locking layer (LL) growth rate on delta layer confinement and electrical properties. (a, b) Measured and fitted SIMS concentration profiles of samples with different LL growth rate. Samples LL-T3 and LL-R1 in (a) do not have a LL RTA. Samples LL-R1-RTA and LL-R2 RTA in (b) have a LL RTA. (c) The reconstructed P concentration profiles. (d) The total and activated P locking probability within 1nm and 2nm from the initial dosing plane as a function of LL thickness. (e) The delta layer free carrier mobility $\mu$ ($cm^2/(V\,s)$) and 2D sheet carrier density $n_s$ ($cm^{-2}$) are characterized at T=2 K using the van der Pauw technique.

Our fitting results show that the LL segregation length decreases with increasing LL growth rate at room temperature. (Figure 7) The segregation length values agree very well with the values reported from previous STM and Auger studies at similar growth rates and temperatures.[80] Physically, this segregation length dependence on growth rate arises from the time allowed for a dopant on the growth front to exchange its lattice position with newly deposited Si atoms before incorporation.[56, 81] Increasing the LL growth rate reduces the time allowed for segregation exchange during LL overgrowth, and therefore increases the incorporation probability within the LL (Table 1) and dopant confinement.

As can be seen in Figure 8 (a, b, c), increasing LL growth rate improves dopant confinement in situations with and without a LL RTA. Increasing the LL growth rate from 0.6ML/min (Sample LL-T3) to 1.1ML/min (Sample LL-R1) increased the P density at the dosing plane from $2.5{\times}10^{21}/cm^3$ to $3.5{\times}10^{21}/cm^3$. At 1.8ML/min LL growth rate in Sample LL-R2-RTA, 95% of the P atoms can be confined within a 1nm thick layer (Figure 8(d)). However, as can be seen from Figure 8(e), which presents both carrier concentration for the four samples as well as Hall mobility, increasing LL growth rate results in decreased dopant activation ratio in samples with and without a 14 second LL RTA at 384°C. While the activated P locking probability decreases, the total P locking probability increases with increased LL growth rate.

Even though P is better confined through either increasing the LL thickness or increasing LL growth rate, we emphasize the advantages of increasing LL growth rate to improve P confinement. As can be seen in Figure 5(b), increasing LL thickness merely extends the P concentration profile within the LL further into the exponential tail while the exponent remains unchanged. While a thicker locking layer can effectively reduce the remaining P coverage on the LL surface that further segregates during the subsequent encapsulation overgrowth at elevated temperature, it has no effect on the P concentration peak height within the LL. On the other hand, increasing the LL growth rate effectively increases the exponent of P profiles within the LL (Figure 8(c)), which improves both the sharpness and concentration peak height of the P profile.

## Discussion

Dopant confinement and electrical activation are highly sensitive to LL fabrication processes at the ML scale. Due to the low segregation probability during LL overgrowth at room temperature, increasing the LL thickness improves delta layer confinement by suppressing the number of dopant atoms that further segregate during the subsequent encapsulation overgrowth at elevated temperature. However, crystalline quality at the LL interface degrades with increased LL thickness which results in lower P activation ratios and free carrier mobilities. Therefore, we identify optimal LL thicknesses that balance dopant confinement and activation at a fixed LL growth rate. In this study, we found such an optimal LL thickness to be approximately 11 ML when depositing the LL at 0.6ML/min, where 90% of P atoms are confined within 2nm of the original dosing plane with an activation ratio of 88%. P density at the original dosing plane is independent of LL thickness, and we estimate a P peak concentration of about $2.5\times10^{21}/cm^3$ can be achieved at a 0.6ML/min LL growth rate.

RTA after LL overgrowth improves both the dopant activation ratio and free carrier mobility. This increase in carrier mobility after a LL RTA occurs because increased Coulomb scattering from additional ionized impurities is offset by decreased point defect scattering due to improved crystal quality, which results in a net increase in the carrier mobility. However, the LL RTA broadens the P distribution within the LL and accumulates P on the LL surface which increases the number of P atoms that segregate during encapsulation layer overgrowth. We note that our calculated P diffusivity within RT-grown LLs at 384℃ is about $3.2\times10^{-17}cm^2/s$, which is approximately three orders of magnitude higher than the corresponding P diffusivity extracted from previous studies within bulk Si at high P concentration ($\sim1.3\times10^{-20}cm^2/s$, following Eq. 2 in Ref. 75).[75, 82] This is likely due to the higher concentration of structural or charge defect complexes within the RT-grown LL and the non-equilibrium local point defect concentration near the highly doped delta layer region and the relatively rough LL surface.[18, 73, 82-84] Elemental SIMS analysis shows a high concentration of atomic point defects due to oxygen, hydrogen and other contaminants that are incorporated into the overgrown locking layers which do not have a significant effect on epitaxy but likely enhance dopant diffusion.[79, 85] We are not aware of any literature values of P diffusivity in the low temperature range of this study and with similar Si LL configurations. Further studies are needed to characterize the detailed physical mechanism(s) of the observed high P diffusivity within RT-grown LLs on Si surfaces with high P coverage.[84]

Increasing the LL growth rate decreases the LL segregation length and improves dopant confinement more efficiently than merely increasing LL thickness in the sense that both the sharpness and peak height of the P concentration profile can be improved within the LL. However, higher LL growth rates affect the local crystal quality at the LL interface and compromise dopant activation.[71] The drop in activated P locking probability (Figure 8(e)) with higher LL growth rates highlights the side effect of improving P confinement by increasing LL growth rate, which can be mitigated by a short LL RTA. Increasing the LL growth rate from 0.6ML/min to 1.1ML/min results in a drop in P activation ratio from 88% to 61% while the mobility increases from $75cm^2/Vs$ to $83cm^2/Vs$. The competing response of activation ratio and free carrier mobility to increased LL growth rate may suggest that the mobility is primarily limited by Coulomb scattering from ionized impurities for room-temperature grown LL without a LL RTA. On the other hand, for LL samples with a LL RTA, an increased LL growth rate results in a reduction of both the P activation ratio and free carrier mobility. Further study is necessary to fully explore the electronic transport dependence on the LL overgrowth parameters. In addition, in order to fully explain the detailed physical mechanisms of P segregation and diffusion in this study, it might be necessary to extend our simple model with additional complexities, such as the growth front roughness,[44] step density[56] evolution, vacuum contamination and auto-dosing,[25, 85, 86] *etc.*, which are beyond the scope of this study.

## Conclusions

To summarize, we have developed a robust quantification method using room-temperature grown locking layers (LL) and segregation/diffusion and sputter profiling simulations to monitor and control, at the atomic scale, unintentional dopant movement and lattice defect formation during the Si:P monolayer fabrication. By combining SIMS, TEM, STM, APT, and low-temperature magnetotransport measurements, it is shown that increasing LL thickness decreases both the dopant activation ratio and carrier mobility. Specific LL growth rates correspond to optimal LL thicknesses that balance the tradeoff between dopant confinement and activation. LL RTA restores LL crystalline quality but induces dopant diffusion and surface accumulation at the LLs. The dopant segregation length can be suppressed below one Si lattice constant by increasing LL growth rate above 1.8ML/min. We compare the effects of increasing LL growth rate and increasing LL thickness on delta layer quality, emphasizing the advantage of the former in improving P confinement in both the profile sharpness and peak concentration heights. We demonstrate that high LL growth rates in combination with a low-temperature LL RTA can create exceptionally sharp dopant confinement while maintaining good electrical quality within Si:P monolayers. The new model developed in this study provides valuable insight into the interplay among dopant movement, activation, and surface roughening at the mono-atomic layer scale. The locking layer fabrication and quantification methods demonstrated in this study provide unique tools to study atomic dopant movement and the local crystalline environment in Si:P monolayers and their effect on atomic scale electronics for future semiconductor and solid state quantum devices.

**CONFLICTS OF INTEREST**

There are no conflicts of interest to declare.

**ACKNOWLEDGMENTS**

This work was sponsored by the Innovations in Measurement Science (IMS) project at NIST: Atom-base Devices: single atom transistors to solid state quantum computing. The authors thank Stephen Smith (Evans Analytical Group) for SIMS measurements and David Simons, Joe Bennett, Joshua Pomeroy, Theodore Einstein, Scott Schmucker, and Ian Appelbaum for useful conversations.